\documentclass[journal]{IEEEtran}
\usepackage{amsmath}
\usepackage{times}
\usepackage{latexsym}
\usepackage{amssymb}
\usepackage{graphicx}
\usepackage{subfigure}
\usepackage{amsfonts}
\usepackage{epsfig}
\begin{document}

\title{\huge Study of Interference Cancellation and Relay Selection Algorithms Using Greedy Techniques for Cooperative DS-CDMA Systems}
\author{Jiaqi~Gu{$^*$}~
        and Rodrigo C. de~Lamare,~\IEEEmembership{Senior Member,~IEEE}

\thanks{J. Gu* is with the Communications Research Group, Department of Electronics, University of York, YO10 5DD York, U.K.
(e-mail: jg849@york.ac.uk).}
\thanks{R. C. de Lamare is with the CETUC, Pontifical Catholic University of Rio de Janeiro (PUC-RIO), Rio de Janeiro, Brazil. He is also with the Communications Research Group, Department of Electronics, University of York, YO10 5DD York, U.K. (e-mail: rodrigo.delamare@york.ac.uk).}
\thanks{This work is funded by the ESII consortium under task 26 for
low-cost wireless ad hoc and sensor networks}
\thanks{Part of this work has been presented at the European Signal Processing Conference (EUSIPCO), 2014 and the International Conference on Acoustics, Speech and Signal Processing (ICASSP), 2014, respectively.}}

\maketitle
\vspace{-6.25em}
\begin{abstract}
In this work, we study interference cancellation techniques and a
multi-relay selection algorithm based on greedy methods for the
uplink of cooperative direct-sequence code-division multiple access
(DS-CDMA) systems. We first devise low-cost list-based successive
interference cancellation (GL-SIC) and parallel interference
cancellation (GL-PIC) algorithms with RAKE receivers as the
front-end that can approach the maximum likelihood detector
performance and be used at both the relays and the destination of
cooperative systems. Unlike prior art, the proposed GL-SIC and
GL-PIC algorithms exploit the Euclidean distance between users of
interest and the potential nearest constellation point with a chosen
threshold in order to build an effective list of detection
candidates. A low-complexity multi-relay selection algorithm based
on greedy techniques that can approach the performance of an
exhaustive search is also proposed. A cross-layer design strategy
that brings together the proposed multiuser detection algorithms and
the greedy relay selection is then developed along with an analysis
of the proposed techniques. Simulations show an excellent bit error
rate performance of the proposed detection and relay selection
algorithms as compared to existing techniques.
\end{abstract}


\IEEEpeerreviewmaketitle

\section{Introduction}
\IEEEPARstart{M}{ultipath} fading is a major constraint that seriously limits the
performance of wireless communications. Indeed, severe fading has a detrimental
effect on the received signals and can lead to a degradation of the
transmission of information and the reliability of the network.
Cooperative diversity is a technique that has been
widely considered in recent years \cite{Proakis} as an effective tool to
deal with this problem. Several cooperative schemes have been proposed in the
literature \cite{sendonaris,Venturino,laneman04}, and among the most
effective ones are Amplify-and-Forward (AF) and
Decode-and-Forward (DF) \cite{laneman04}. For an AF protocol,
relays cooperate and amplify the received signals with a given
transmit power amplifying their own noise. With the DF protocol, relays decode the received
signals and then forward the re-encoded message to the destination.
Consequently, better performance and lower power consumption can be
obtained when appropriate decoding and relay selection strategies
are applied.

\subsection{Prior and related work}
DS-CDMA systems are a multiple access technique that can be
incorporated with cooperative systems in ad hoc and sensor networks
\cite{Bai,Souryal,Levorato}. Due to the multiple access
interference (MAI) effect that arises from nonorthogonal received waveforms and narrowband interfering signals,
the system performance may be adversely affected. To deal with this issue,
multiuser detection (MUD) techniques have been developed \cite{Verdu1} as an effective approach
to suppress MAI. The optimal detector, known as maximum likelihood (ML)
detector, has been proposed in \cite{Verdu2}. However, this
method is infeasible for ad hoc and sensor networks considering its
computational complexity. Motivated by this fact, several
sub-optimal strategies have been developed: the linear detector
\cite{Lupas}, the successive interference cancellation (SIC)
\cite{Patel}, the parallel interference
cancellation (PIC) \cite{Varanasi} and the minimum mean-square error
(MMSE) decision feedback detector \cite{RCDL1}.
A key challenge is how to design interference cancellation techniques with low cost and near ML performance.
Moreover, such interference cancellation algorithms should be suitable to cooperative relaying systems
and feasible for deployment at the relays and small devices.

In cooperative relaying systems, different strategies that utilize
multiple relays have been recently introduced in
\cite{Jing,Clarke,Ding,Song,Talwar}. Among these approaches, a
greedy algorithm is an effective way to approach the global optimal
solution. Greedy algorithms are important mathematical techniques
that follow the approach of obtaining a locally optimal solution to
complex problems with low cost in a step by step manner. Decisions
at each step in the greedy process are made to provide the largest
benefit based on improving the local state without considering the
global situation. Greedy algorithms may fail to achieve the globally
optimal choice as they do not execute all procedures exhaustively,
however, they are still useful as they usually present a lower cost
and can provide acceptable approximations. Greedy algorithms have
been widely applied in sparse approximation \cite{Tropp}, internet
routing \cite{Flury} and arithmetic coding \cite{Jia}. In
\cite{Tropp}, orthogonal matching pursuit (OMP) and basis pursuit
(BP) are two major greedy approaches that are used to approximate an
arbitrary input signal with the near optimal linear combination of
various elements from a redundant dictionary. In \cite{Flury},
greedy routing is mentioned as a routing strategy where messages are
simply forwarded to the node that is closest to the destination. In
order to reduce the computational complexity and improve the overall
speed of arithmetic coding, a greedy re-normalization step that
contains greedy thresholding and greedy outputting is proposed and
analyzed in \cite{Jia}. In relay-assisted systems, greedy algorithms
are used in \cite{Ding,Song} to search for the best set of relays,
however, with insufficient numbers of combinations considered, a
significant performance loss is experienced as compared to an
exhaustive search.

\subsection{Contributions}
This work presents cost-effective interference cancellation
algorithms and multi-relay selection algorithms for cooperative
DS-CDMA systems. The proposed interference cancellation algorithms
do not require matrix inversions and rely on the RAKE receiver as
the front-end. A cross-layer optimization approach that jointly
considers the proposed interference cancellation and relay selection
algorithms for ad hoc and sensor networks is also proposed. The
proposed techniques are not limited to DS-CDMA systems and could
also be applied to multi-antenna and multi-carrier systems.
Cross-layer designs that integrate different layers of the network
have been employed in prior work \cite{RCDL2,Chen} to guarantee the
quality of service and help increase the capacity, reliability and
coverage of systems. However, MUD techniques with relay selection in
cooperative relaying systems have not been discussed widely in the
literature. In \cite{Venturino,Cao}, an MMSE-MUD technique has been
applied to cooperative systems, where the results indicate that the
transmissions are more resistant to MAI and obtain a significant
performance gain when compared with a single direct transmission.
However, extra costs are introduced, as matrix inversions are
required when an MMSE filter is deployed.

The contributions of this paper are summarized as follows:
\begin{itemize}
\item We propose a low-cost greedy list-based successive interference cancellation (GL-SIC) multiuser detection method that can be applied at both the
      relays and the destination of wireless systems. 
\item We also develop a low-cost greedy list-based parallel interference cancellation (GL-PIC) strategy which employs RAKE receivers as the front-end
      and can approach the ML detector performance. 
\item We present a low-complexity multi-relay selection algorithm based on greedy techniques that can approach the performance of an exhaustive
      search. 
\item An analysis of the computational complexity, the greedy relay selection method and the cross-layer design is presented.
\item A cross-layer design that incorporates the optimization of the proposed GL-SIC and GL-PIC techniques and the improved greedy multi-relay selection algorithm for the
      uplink of a cooperative DS-CDMA system is developed and evaluated. 
\end{itemize}

The rest of this paper is organized as follows. In Section II, the
system model is described. In Section III, the GL-SIC multiuser
detection method is presented. In Section IV, the GL-PIC multiuser
detection method is then developed. In Section V, the relay selection
strategy is proposed. In Section VI, the computational complexity and the greedy relay selection process are analyzed.
In Section VII, the cross-layer design is explained. In Section VIII, simulation results are
presented and discussed. Finally, conclusions are drawn in Section
IX.

Notation: in this paper, we use boldface upper and boldface lower fonts to denote matrices and vectors, respectively. $(.)^T$ and $(.)^H$ represent the transpose and Hermitian transpose, respectively. $(.)^{-1}$ stands for the matrix inversion, $E[.]$ denotes the expected value, $\mid.\mid$ indicates the norm of a scalar and $\parallel.\parallel$ implies the norm of a vector.

\section{Cooperative DS-CDMA system model}

\begin{figure}[!htb]
\begin{center}
\def\epsfsize#1#2{0.45\columnwidth}
\epsfbox{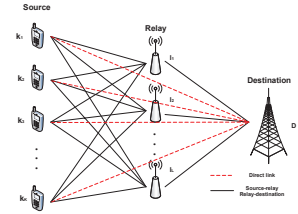} \caption{\footnotesize
Uplink of a cooperative DS-CDMA system.}  \label{fig1}
\end{center}
\vspace{-1em}
\end{figure}

We consider the uplink of a synchronous DS-CDMA system with $K$
users $(k_1,k_2,...k_K)$, $L$ relays $(l_1,l_2,...l_L)$, $N$ chips per
symbol and $L_p$ $(L_p<N)$ propagation paths for each link. The system
is equipped with a DF protocol at each relay and we assume that the
transmit data are organized in packets comprising $P$ symbols. The
received signals are filtered by a matched filter, sampled at chip
rate to obtain sufficient statistics and organized into $M \times1$
vectors $\textbf{y}_{sd}$, $\textbf{y}_{sr}$ and $\textbf{y}_{rd}$,
which represent the signals received from the sources (users) to the
destination, the sources to the relays and the relays to the
destination, respectively. The proposed algorithms for interference
mitigation and relay selection are employed at the relays and at the
destination. As shown in Fig.\ref{fig1}, the cooperation takes
place in two phases. During the first phase, the received data at the destination
and the $l$-th relay can be described by
\begin{equation}
\textbf{y}_{sd}= \sum\limits_{k=1}^Ka_{sd}^k\textbf{S}_k
\textbf{h}_{sd,k}b_k+\textbf{n}_{sd},
\end{equation}
\begin{equation}
 \textbf{y}_{sr_l}=\sum\limits_{k=1}^Ka_{sr_l}^k\textbf{S}_k\textbf{h}_{sr_l,k}b_k+\textbf{n}_{sr_l},
\end{equation}

where $M=N+L_p-1$, $b_k\in\{+1,-1\}$ correspond to the transmitted
symbols, $a_{sd}^k$ and $a_{sr_l}^k$ represent the $k$-th user's amplitude from
the source to the destination and the source to relay $l$. The vectors
$\textbf{h}_{sd,k}$, $\textbf{h}_{sr_l,k}$ are the $L_p\times1$
channel vectors for user $k$ from the source to the destination and the source
to relay $l$, respectively. The $M\times1$ noise vectors
$\textbf{n}_{sd}$ and $\textbf{n}_{sr_l}$ contain samples of zero
mean complex Gaussian noise with variance $\sigma^2$. The $M \times
L_p$ matrix $\textbf{S}_k$ contains the signature sequence of each
user shifted down by one position at each column that forms
\begin{equation}
\textbf{S}_k = \left[\begin{array}{c c c }
s_{k}(1) &  & {\bf 0} \\
\vdots & \ddots & s_{k}(1)  \\
s_{k}(N) &  & \vdots \\
{\bf 0} & \ddots & s_{k}(N)  \\
 \end{array}\right],
\end{equation}
where $\textbf{s}_k=[s_k(1),s_k(2),...s_k(N)]^T$ is the signature
sequence for user $k$. During the second phase of the transmission,
each relay decodes and reconstructs the received signals using a DF
protocol, then they forward the processed signals to the
destination. It is assumed that each relay  is perfectly
synchronized and transmits at the same time, the signals received at
the destination are then expressed by
\begin{equation}
\textbf{y}_{rd}= \sum\limits_{l=1}^{L}
\sum\limits_{k=1}^K a_{r_ld}^k\textbf{S}_k
\textbf{h}_{r_ld,k}\hat{b}_{r_ld,k}+\textbf{n}_{rd}, \label{equation0}
\end{equation}
where $a_{r_ld}^k$ is the amplitude for source (user) $k$ from the
$l$-th relay to the destination, $\textbf{h}_{r_ld,k}$ is the $L_p
\times 1$ channel vector for user $k$ from the $l$-th relay to the destination,
$\textbf{n}_{rd}$ is the $M\times1$ zero mean complex Gaussian noise with
variance $\sigma^2$, $\hat{b}_{r_ld,k}$ is the decoded symbol at the
output of relay $l$ after using the DF protocol.

The received signal at the destination comprises the data transmitted during two phases that are
jointly processed at the destination. Therefore, the received signal is described by a
$2M\times1$ vector formed by stacking the received signals
from the relays and the sources as given by
\begin{equation}
\begin{split}
\hspace{-0.5em} \left[\hspace{-0.5em}\begin{array}{l}
  \textbf{y}_{sd}\\
  \textbf{y}_{rd}\\
\end{array} \hspace{-0.5em} \right] & = \left[\hspace{-0.5em} \begin{array}{l}
  \sum\limits_{k=1}^K  a_{sd}^k\textbf{S}_k\textbf{h}_{sd,k}b_k \\
  \sum\limits_{l=1}^{L}\sum\limits_{k=1}^K a_{r_ld}^k\textbf{S}_k\textbf{h}_{r_ld,k}\hat{b}_{r_ld,k}\\
\end{array}\hspace{-0.5em} \right] + \left[ \hspace{-0.5em} \begin{array}{l}
  \textbf{n}_{sd}\\
  \textbf{n}_{rd}\\
  \end{array}\right]. \label{equation1}
\end{split}
\end{equation}
The received signal in (\ref{equation1}) can then be described by
\begin{equation}
\textbf{y}_d(i)=\sum\limits_{k=1}^K \textbf{C}_k
\textbf{H}_k(i)\textbf{A}_k(i)\textbf{B}_k(i)+\textbf{n}(i),
\end{equation}
where $i$ denotes the time instant corresponding to one symbol in the
transmitted packet and its received and relayed copies. $\textbf{C}_k$
is a $2M\times(L+1)L_p$ matrix comprising shifted versions of $\textbf{S}_k$ as given by
\begin{equation}
\textbf{C}_k = \left[\begin{array}{c c c c}
\textbf{S}_{k} & {\bf 0} & \ldots & {\bf 0} \\
{\bf 0} & \ \ \textbf{S}_{k} & \ldots & \ \ \textbf{S}_{k}\\
 \end{array}\right],
\end{equation}
$\textbf{H}_k(i)$ represents a $(L+1)L_p \times (L+1)$ channel matrix
between the sources and the destination and the relays and the
destination links. $\textbf{A}_k(i)$ is a $(L+1)\times(L+1)$
diagonal matrix of amplitudes for user $k$.
$\textbf{B}_k(i)=[b_k,\hat{b}_{r_1d,k},\hat{b}_{r_2d,k},...\hat{b}_{r_Ld,k}]^T$
is a $(L+1)\times1$ vector for user $k$ that contains the transmitted symbol at
the source and the detected symbols at the output of each relay, and
$\textbf{n}(i)$ is a $2M\times1$ noise vector.

\section{Proposed GL-SIC multiuser detection}
In this section, we detail the GL-SIC multiuser detector that can be
applied in the uplink of a cooperative system. The GL-SIC detector
uses the RAKE receiver as the front-end, so that the matrix
inversion required by the MMSE filter can be avoided
\cite{int,Chen,Meng,l1cg,zhaocheng,alt,jiolms,jiols,jiomimo,jidf,fa10,saabf,barc,honig,mswfccm,song,locsme}.
The GL-SIC detector exploits the Euclidean distance between the
users of interest and their nearest constellation points, with
multiple ordering at each stage, all possible lists of tentative
decisions for each user are generated. When seeking appropriate
candidates, a greedy-like technique is performed to build each list
and all possible estimates within the list are examined when
unreliable users are detected. Unlike prior work which employs the
concept of Euclidean distance with multiple feedback SIC (MF-SIC)
\cite{Li1}, GL-SIC does not require matrix inversions and jointly
considers multiple numbers of users, constellation constraints and
re-ordering at each detection stage to obtain an improvement in
detection performance.

\subsection{Proposed GL-SIC design}

\begin{figure}[!htb]
\begin{center}
\def\epsfsize#1#2{0.65\columnwidth}
\epsfbox{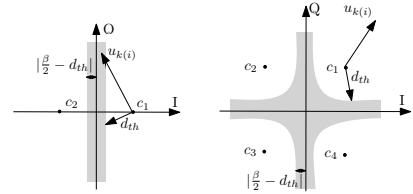} \caption{The reliability check in BPSK and QPSK
constellations.} \label{fig2}
\end{center}
\end{figure}

In the following, we describe the process for initially detecting
$n$ users described by the indices $k_1, k_2,...,k_n$ at the first
stage. Other users can be obtained accordingly. As shown by Fig.
\ref{fig2}, $\beta$ is the distance between two nearest
constellation points, $d_{th}$ is the threshold. The soft output of
the RAKE receiver for user $k$ is then obtained by
\begin{equation}
u_{k}(i)=\textbf{w}_{k}^{H}\textbf{y}_{sr_l}(i),
\end{equation}
where $\textbf{y}_{sr_l}(i)$ represents the received signal from the
source to the $l$-th relay, $u_{k}(i)$ stands for the soft output of
the $i$-th symbol for user $k$ and $\textbf{w}_{k}$ denotes the RAKE
receiver that corresponds to a filter matched to the effective
signature at the receiver. After that, we order all users according
to a decreasing power level and organize them into a $K \times 1$
vector $\textbf{t}_a$. We choose the first $n$ entries
$[\textbf{t}_a(1), \textbf{t}_a(2),...,\textbf{t}_a(n)]$ which
denote users $k_1, k_2,...,k_n$, the reliability of each of the $n$
users is examined by the corresponding Euclidean distance between
the desired user and its nearest constellation point $c$ as
explained next.
\\
\\
\textbf{Decision reliable}:\\
If all $n$ users are considered reliable
\begin{equation}
u_{\textbf{t}_a(t)}(i)\notin \textbf{C}_{\rm grey},\ \ \ {\rm for} \ t\in [1,2,...,n],
\end{equation}
these soft estimates will then be applied to a slicer $Q(\cdot)$ as
\begin{equation}
\hat{b}_{\textbf{t}_a(t)}(i)=Q(u_{\textbf{t}_a(t)}(i)),\ {\rm for}\ t\in [1,2,...,n],
\end{equation}
where $\hat{b}_{\textbf{t}_a(t)}(i)$ denotes the detected symbol for the $\textbf{t}_a(t)$-th user, $\textbf{C}_{\rm grey}$ is the shadowed area in Fig. \ref{fig2}, it should be noted that the shadowed region would spread along both the vertical and horizontal directions. The cancellation is then performed in the same way as a conventional SIC where we mitigate the impact of MAI brought by these users
\begin{equation}
\textbf{y}_{sr_l,s+1}(i)=\textbf{y}_{sr_l,s}(i)-\sum\limits_{t=1}^{n}\textbf{H}_{sr_l,\textbf{t}_a(t)}(i)
\hat{b}_{\textbf{t}_a(t)}(i), \label{equation2}
\end{equation}
where $\textbf{H}_{sr_l,\textbf{t}_a(t)}(i) = a_{sr_l}^{\textbf{t}_a(t)}\textbf{S}_{\textbf{t}_a(t)}(i)\textbf{h}_{sr_l,\textbf{t}_a(t)}(i)$
stands for the desired user's channel matrix associated with the link between the
source and the $l$-th relay, $\textbf{y}_{sr_l,s}$ is the received
signal from the source to the $l$-th relay at the $s$-th $(s=1,2,...,K/n)$ cancellation stage.
The process is then repeated with another $n$ users being selected from the remaining users at each following stage,
and this algorithm changes to the unreliable mode when unreliable users are detected. Additionally, since the interference
created by the previous users with the strongest power has been mitigated, improved estimates are obtained by reordering the remaining users.
\\
\\
\textbf{Decision unreliable}:\\
(a). If part of the $n$ users are determined as reliable, while others
are considered as unreliable, we have
\begin{equation}
u_{\textbf{t}_p(t)}(i) \notin \textbf{C}_{\rm grey},\ \ \ {\rm for}\ t\in [1,2,...,n_p], \label{equation3}
\end{equation}
\begin{equation}
u_{\textbf{t}_q(t)}(i) \in \textbf{C}_{\rm grey},\ \ \ {\rm for}\ t\in [1,2,...,n_q], \label{equation4}
\end{equation}
where $\textbf{t}_p$ is a $1 \times n_p$ vector that
contains $n_p$ reliable users and $\textbf{t}_q$ is a $1 \times n_q$ vector
that includes $n_q$ unreliable users, subject to $\textbf{t}_p\cap \textbf{t}_q=\varnothing$ and $ \textbf{t}_p\cup \textbf{t}_q=[1,2,...n]$ with $n_p+n_q=n$. Consequently, the $n_p$ reliable users are applied to the slicer $Q(\cdot)$ directly and the $n_q$ unreliable ones are examined in terms of all possible constellation values $c^m$ $(m=1,2,...,N_c)$ from the $1\times N_c$ constellation points set $\textbf{C}\subseteq \textsl{F}$, where $\textsl{F}$ is a subset of the complex field and $N_c$ is determined by the modulation type. The detected symbols are given by
\begin{equation}
\hat{b}_{\textbf{t}_p(t)}(i)=Q(u_{\textbf{t}_p(t)}(i)),
{\rm for}\ t\in [1,2,...,n_p],
\end{equation}
\begin{equation}
\hat{b}_{\textbf{t}_q(t)}(i)=c^m,\ \
{\rm for}\ t\in [1,2,...,n_q],
\end{equation}
At this point, $N_c^{n_q}$ combinations of candidates for $n_q$ users are generated. The detection tree is then split into $N_c^{n_q}$ branches. After this processing, (\ref{equation2}) is applied with its corresponding combination to ensure the interference caused by the $n$ detected users is mitigated. Following that, $N_c^{n_q}$ numbers of updated $\textbf{y}_{sr_l}(i)$ are generated, we reorder the remaining users at each cancellation stage and compute a conventional SIC with RAKE receivers for each branch.

The following $K\times1$ different ordered candidate detection lists
are then produced:
\begin{equation}
\textbf{b}^j(i)=[\textbf{s}_{\textrm{pre}}(i), \ \ \textbf{s}_p(i),\ \ \textbf{s}^j_q(i),\ \ \textbf{s}^j_{\textrm{next}}(i)]^T, \ j=1,2,...,N_c^{n_q},
\end{equation}
where
\begin{itemize}
\item[] $\textbf{s}_{\textrm{pre}}(i)=[\hat{b}_{\textbf{t}_a(1)}(i),\hat{b}_{\textbf{t}_a(2)}(i),...]^T$ stands for the previous stages detected reliable
        symbols,
\item[] $\textbf{s}_p(i)=[\hat{b}_{\textbf{t}_p(1)}(i), \hat{b}_{\textbf{t}_p(2)}(i),...,\hat{b}_{\textbf{t}_p(n_p)}(i)]^T$ is a $n_p\times 1$ vector that
        denotes the current stage reliable symbols detected directly from slicer $Q(\cdot)$ when (\ref{equation3}) occurs,
\item[] $\textbf{s}^j_q(i)=[c_{\textbf{t}_q(1)}^{m},c_{\textbf{t}_q(2)}^{m},...,c_{\textbf{t}_q(n_q)}^{m}]^T, j=1,2,...,N_c^{n_q}$ is a $n_q\times 1$ vector that contains the detected symbols deemed unreliable at the current stage as in (\ref{equation4}), each entry of this vector is selected randomly from the constellation point set $\textbf{C}$ and all possible $N_c^{n_q}$ combinations need to be considered and examined.
\item[]  $\textbf{s}^j_{\textrm{next}}(i)=[...,\hat{b}_{\textbf{t}'(1)}^{\textbf{s}^j_q}(i),...,\hat{b}_{\textbf{t}'(n)}^{\textbf{s}^j_q}(i)]^T$ includes
         the corresponding detected symbols in the following stages after the $j$-th combination of $\textbf{s}_q(i)$ is allocated to the unreliable user vector $\textbf{t}_q$,
\item[]  $\textbf{t}'$ is a $n \times 1$ vector that contains the users from the last stage.
\end{itemize}

(b). If all $n$ users are considered as unreliable, then we have
\begin{equation}
u_{\textbf{t}_b(t)}(i) \in \textbf{C}_{\rm grey},\ \ \ {\rm for}\ t\in [1,2,...,n],\label{equation5}
\end{equation}
where $\textbf{t}_b=[1,2,...,n]$ and all $n$ unreliable users can
assume the values in $\textbf{C}$. In this case, the detection tree
will be split into $N_c^{n}$ branches to produce
\begin{equation}
\hat{b}_{\textbf{t}_b(t)}(i)=c^m, \ {\rm for}\ t\in [1,2,...,n],
\end{equation}
Similarly, (\ref{equation2}) is then applied and a conventional SIC with different orderings at each cancellation stage
is performed via each branch.

Since all possible constellation values are tested for all unreliable users, we have the candidate lists
\begin{equation}
\textbf{b}^j(i)=[\textbf{s}_{\textrm{pre}}(i),\ \textbf{s}_b^j(i),\ \ \textbf{s}^j_{\textrm{next}}(i)]^T, \ j=1,2,...,N_c^n,
\end{equation}
where
\begin{itemize}
\item[] $\textbf{s}_{\textrm{pre}}(i)=[\hat{b}_{\textbf{t}_a(1)}(i),\hat{b}_{\textbf{t}_a(2)}(i),...]^T$ are the reliable symbols that are detected from
         previous stages,
\item[] $\textbf{s}_b^j(i)=[c_{\textbf{t}_b(1)}^{m},c_{\textbf{t}_b(2)}^{m},...,c_{\textbf{t}_b(n)}^{m}]^T, j=1,2,...,N_c^n$ is a $n\times1$ vector that
        represents the number of users $n$ which are regarded as unreliable at the current stage as shown by (\ref{equation5}), each entry of $\textbf{s}_b^j$ is selected randomly from the constellation point set $\textbf{C}$.
\item[] The vector $\textbf{s}^j_{\textrm{next}}(i)=[...,\hat{b}_{\textbf{t}'(1)}^{\textbf{s}_b^j}(i),...,\hat{b}_{\textbf{t}'(n)}^{\textbf{s}_b^j}(i)]^T$
        contains the corresponding detected symbols in the following stages after the $j$-th combination of $\textbf{s}_b(i)$ is allocated to all unreliable users.
\end{itemize}
After the candidates are generated, lists are built for each group of users, and the ML rule is used 
to choose the best candidate list as described by
\begin{equation}
\textbf{b}^{\textrm{best}}(i)= \min _{\substack{1\leq j\leq m, \textrm{where}\\m=N_c^{n_q} \textrm{or}\ N_c^n}}\parallel \textbf{y}_{sr_l}(i)-\textbf{H}_{sr_l}(i)\textbf{b}^{j}(i)\parallel^2.
\end{equation}
The proposed GL-SIC algorithm is detailed in Table I.
\begin{table}[!htb]\scriptsize
\centering\caption{The GL-SIC algorithm}
\begin{tabular}{l}
\hline
$u_{k}(i)=\textbf{w}_{k}^{H}\textbf{y}_{sr_l}(i)$ \% soft outputs of all candidates\\
order $u_{k}(i)$ according to a decreasing power level and organize them \\
into \textbf{$\textbf{t}_a$}\\
\textbf{for} s = 1 \textbf{to} $K/n$\\\ \ \ \ \ \textbf{if} no
unreliable users have been detected\\ \ \ \ \ \ \ \ \ \textbf{for}
t=1: n\\\ \ \ \ \ \ \ \ \ \ \ \ \textbf{if}
$u_{\textbf{t}_a(t)}(i)\notin \textbf{C}_{\rm grey}$ \%
\textbf{reliable}\\ \ \ \ \ \ \ \ \ \ \ \ \ \ \ \
$\hat{b}_{\textbf{t}_a(t)}(i)=Q(u_{\textbf{t}_a(t)}(i))$\\\ \ \ \ \
\ \ \ \ \ \ \textbf{else}\ \ \ \ \% \textbf{unreliable}\\\ \ \ \ \ \
\ \ \ \ \ \ \ \ \ $\hat{b}_{\textbf{t}_a(t)}(i)=c^m$\\\ \ \ \  \ \ \
\ \ \ \ \textbf{end}\\ \ \ \ \ \ \ \ \ \textbf{end}\\ \ \ \ \ \ \ \
\ \textbf{Do conventional SIC via each branch}\\ \ \ \ \
\textbf{else} \% unreliable users have already been detected at
previous stages\\ \ \ \ \ \ \ \ \ \ \textbf{Re-order the n soft
estimates for stage s and send them}\\ \ \ \ \ \ \ \ \ \ \textbf{to
the slicer $Q(\cdot)$}\\ \ \ \ \ \ \ \ \ \ \textbf{Perform
conventional SIC in each branch}\\ \ \ \ \
\textbf{end}\\
\textbf{end}\\
\% apply ML to choose the best candidates list\\
$\textbf{b}^{\textrm{best}}(i)= \min _{\substack{1\leq j\leq m, \textrm{where}\\m=N_c^{n_q} \textrm{or}\ N_c^n}}\parallel \textbf{y}_{sr_l}(i)-\textbf{H}_{sr_l}(i)\textbf{b}^{j}(i)\parallel^2$\\
\hline
\vspace{-5.85em}
\end{tabular}
\end{table}

\subsection{GL-SIC with multi-branch processing}

The multiple branch (MB) structure \cite{RCDL1,Li2} that employs
multiple parallel processing branches can help to obtain extra
detection diversity. Inspired by the MB approach \cite{RCDL1,Li2},
we change the obtained best detection order for
$\textbf{b}^{\textrm{best}}$ with indices $\textbf{O}=[1,2,...,K]$
into a group of different detection sequences to form a parallel
structure with each branch shares a different detection order. This
approach generates lists with further candidates for detection and
can further improve the performance of GL-SIC. Since it is not
practical to test all $L_b=K!$ possibilities due to the high
complexity, a reduced number of branches is employed. Note that a
small number of branches captures most of the performance gains and
allow the GL-SIC with the MB technique to approach the ML
performance. With each index number in $\textbf{O}_{l_b}$ being the
corresponding index number in $\textbf{O}$ cyclically shifted to the
right by one position as shown by
\begin{itemize}
\item[]  $\textbf{O}_{l_1}=[K,1,2,...,K-2,K-1]$,
\item[]  $\textbf{O}_{l_2}=[K-1,K,1,...,K-3,K-2]$,
\item[]  \ \ \ \ \ \ \ \ \ \ \ \ \ \ \ \ \ $\vdots$
\item[]  $\textbf{O}_{l_{K-1}}=[2,3,4,...,K,1]$,
\item[]  $\textbf{O}_{l_K}=[K,K-1,K-2,...,2,1]$(reverse order).
\end{itemize}
After that, each of the $K$ parallel branches computes a GL-SIC
algorithm with its corresponding order. After obtaining $K+1$
different candidate lists according to each branch, a modified ML
rule is applied with the following steps:
\begin{enumerate}
\item Obtain the best candidate branch $\textbf{b}^{O_{l_{\rm base}}}(i)$ among all $K+1$ ($\textbf{O}$ included) parallel branches according to the ML
      rule:
      \begin{equation} \label{equation6}
      \textbf{b}^{O_{l_{\rm base}}}(i)= \min _{\substack{0\leq b\leq K}}\parallel \textbf{y}_{sr_l}(i)-\textbf{H}_{sr_l}\textbf{b}^{O_{l_b}}(i)\parallel^2
      \end{equation}
\item Re-examine the detected symbol for user $k$ $(k=1,2,...,K)$ by fixing the detected results of all other unexamined users in
      $\textbf{b}^{O_{l_{\rm base}}}(i)$.

\item Replace the $k$-th user's detection result $\hat b_{k}$ in $\textbf{b}^{O_{l_{\rm base}}}(i)$ by its corresponding detected values from all other $K$
      branches $\textbf{b}^{O_{l_b}}(i)$, $(l_b\neq l_{\rm base}, \textbf{O}=\textbf{O}_{l_0})$ with the same index, the combination with the minimum Euclidean distance is selected through the ML rule and the improved estimate of user $k$ is saved and kept.

\item The same process is then repeated with the next user in $\textbf{b}^{O_{l_{\rm base}}}(i)$ until all users in $\textbf{b}^{O_{l_{\rm base}}}(i)$ are
      examined.
\end{enumerate}

%
The proposed modified ML selection technique is shown in Table II.
\begin{table}[!htb]\scriptsize
\centering\caption{The modified ML selection process}
\begin{tabular}{l}
\hline
$\textbf{b}^{\rm opt}=[]$ \% define an empty vector initially\\
\textbf{for} k = 1 \textbf{to} $K$ \\\ \ \ \
\textbf{for} n = 1 \textbf{to} $K$ \\ \ \ \ \ \ \ \ \ \
$\textbf{b}_{\rm temp}^{\textbf{O}_{l_n}}=[\textbf{b}^{\rm opt},\textbf{b}^{\textbf{O}_{l_n}}[k],\textbf{b}^{\textbf{O}_{l_{\rm base}}}[k+1],...,\textbf{b}^{\textbf{O}_{l_{\rm base}}}[K]]$\\ \ \ \ \
\textbf{end}\\ \ \ \ \
\textbf{Apply ML rule to choose the best combination}\\ \ \ \ \
\% save the corresponding estimate for user k from the selected\\ \ \ \ \ \ \ \
branch $\textbf{O}_{l_{\rm selected}}$ that provides the best combination\\ \ \ \ \
$\textbf{b}^{\rm opt}=[\textbf{b}^{\rm opt}, \textbf{b}^{\textbf{O}_{l_{\rm selected}}}[k]]$\\
\textbf{end}\\
\hline
\vspace{-5.85em}
\end{tabular}
\end{table}

\section{Proposed GL-PIC multiuser detection}

In this section, we present a GL-PIC detector
that can be applied at both the relays and destination in the
uplink of a cooperative system. The GL-PIC detector uses the
RAKE receiver as the front-end, so that the matrix inversion brought by the MMSE filter can be avoided.
Specifically, the proposed GL-PIC algorithm determines the
reliability of the detected symbol by comparing the Euclidean
distance between the symbols of users of interest and the potential
nearest constellation point with a chosen threshold. After checking
the reliability of the symbol estimates by listing all possible
combinations of tentative decisions, the $n_q$ most unreliable users
are re-examined via a number of selected constellation points in a
greedy-like approach, which saves computational complexity by
avoiding redundant processing with reliable users. The soft
estimates of the RAKE receiver for each user are obtained by
\begin{equation}
u_{k}(i)=\textbf{w}_{k}^{H}\textbf{y}_{sr_l}(i),
\end{equation}
As shown in Fig.\ref{fig2},
for the $k$-th user, the reliability of its soft estimates is determined by the
Euclidean distance between $u_{k}(i)$ and its nearest constellation
points $c$.\\
\\
\textbf{Decision reliable}:\\
If the soft estimates of $n_a$ users satisfy the following condition
\begin{equation}
u_{\textbf{t}_a(t)}(i)\notin \textbf{C}_{\rm grey},\ \ \ {\rm for} \ t\in [1,2,...,n_a],
\end{equation}
where $\textbf{t}_a$ is a $1\times n_a$ vector that contains $n_a$ reliable estimates, $\textbf{C}_{\rm grey}$ is the grey area in Fig.\ref{fig2}
and the grey area would extend along both the vertical and horizontal directions. These soft estimates are applied to a slicer $Q(\cdot)$ as
described by
\begin{equation}
\hat{b}_{\textbf{t}_a(t)}(i)=Q(u_{\textbf{t}_a(t)}(i)),\ {\rm for}\ t\in [1,2,...,n_a],
\end{equation}
where $\hat{b}_{\textbf{t}_a(t)}(i)$ denotes the detected symbol for the $\textbf{t}_a(t)$-th user.\\
\\
\textbf{Decision unreliable}:\\
In case that $n_b$ users are determined as unreliable, a $1\times n_b$
vector $\textbf{t}_b$ with $n_b$ unreliable estimates included is produced, as given by
\begin{equation}
u_{\textbf{t}_b(t)}(i)\in \textbf{C}_{\rm grey},\ \ \ {\rm for} \ t\in [1,2,...,n_b],
\end{equation}
we then sort these unreliable estimates in terms of their Euclidean
distance in a descending order. Consequently, the first $n_q$ users from the ordered set are deemed as the most
unreliable ones as they experience the farthest distance to their reference constellation
points. These $n_q$ estimates are then examined in terms of all possible constellation values
$c^m$ $(m=1,2,...,N_c)$ from the $1\times N_c$ constellation points set $\textbf{C}\subseteq \textsl{F}$, where $\textsl{F}$ is a subset of the complex field, and $N_c$ is determined by the modulation type. Meanwhile, the remaining $n_p=n_b-n_q$ unreliable
users are applied to the slicer $Q(\cdot)$ directly, as described by
\begin{equation}
\hat{b}_{\textbf{t}_p(t)}(i)=Q(u_{\textbf{t}_p(t)}(i)),
\ \ {\rm for} \ t\in [1,2,...,n_p],
\end{equation}
\begin{equation}
\hat{b}_{\textbf{t}_q(t)}(i)=c^m,\ \
{\rm for} \ t\in [1,2,...,n_q],
\end{equation}
where $\textbf{t}_p\cap \textbf{t}_q=\varnothing$ and $\textbf{t}_p \cup \textbf{t}_q=\textbf{t}_b$.

Therefore, by listing all possible combinations of elements across the $n_q$ most unreliable
users, the following $K\times1$ tentative candidate decision lists are generated
\begin{equation}
\textbf{b}^j=[\textbf{s}_{a},\ \ \textbf{s}_{p},\ \ \textbf{s}^j_{q}]^T, \ j=1,2,...,N_c^{n_q},
\end{equation}
where
\begin{itemize}
\item[] $\textbf{s}_{a} \ = \ [\hat{b}_{\textbf{t}_a(1)},\hat{b}_{\textbf{t}_a(2)},...,\hat{b}_{\textbf{t}_a(n_a)}]^T$ is a $n_a \times 1$ vector that
         contains the detected values for the $n_a$ reliable users,
\item[] $\textbf{s}_{p}=[\hat{b}_{\textbf{t}_p(1)},\hat{b}_{\textbf{t}_p(2)},...,\hat{b}_{\textbf{t}_p(n_p)}]^T$ is a $n_p \times 1$ vector that represents
         $n_p$ unreliable users that are detected by the slicer $Q(\cdot)$ directly,
\item[] $\textbf{s}^j_q=[c_{\textbf{t}_q(1)}^{m},c_{\textbf{t}_q(2)}^{m},...,c_{\textbf{t}_q(n_q)}^{m}]^T$ is a $n_q \times 1$ tentative candidate
         combination vector. Each entry of the vector is selected randomly from the constellation point set $\textbf{C}$ and all possible $N_c^{n_q}$ combinations need to be considered and examined.
\end{itemize}
The trade-off between performance and complexity is highly related to the
modulation type and the number ($n_q$) of users we choose from $\textbf{t}_b$. Additionally,
the threshold we set at the initial stage is also a key factor that could affect the quality of detection.

After the $N_c^{n_q}$ candidate lists are generated, the ML rule is used subsequently to choose the best candidate list
as described by
\begin{equation}
\textbf{b}^{\textrm{opt}}= \min _{\substack{1\leq j\leq N_c^{n_q}}}\parallel \textbf{y}_{sr_l}(i)-\textbf{H}_{sr_l}\textbf{b}^{j}(i)\parallel^2.
\end{equation}
Following that, $\textbf{b}^{\textrm{opt}}$ is used as the input for a multi-iteration PIC process as described by
\begin{equation}
\hat{b}_k^i=Q(\textbf{H}_{sr_l,k}^H\textbf{y}_{sr_l}-\sum\limits_{\substack{j=1\\j\neq k}}^{K}\textbf{H}_{sr_l,k}^H\textbf{H}_{sr_l,j}\hat{b}_j^{i-1}),
\end{equation}
where $\hat{b}_k^i$ denotes the detected value for user $k$ at the
$i$-th PIC iteration, $\textbf{H}_{sr_l,k}$ and
$\textbf{H}_{sr_l,j}$ stand for the channel matrices for the $k$-th
and $j$-th user from the source to the $l$-th relay, respectively.
$\hat{b}_j^{i-1}$ is the detected value for user $j$ that comes from
the $(i-1)$-th PIC iteration. Normally, the conventional PIC is
performed in a multi-iteration way, where for each iteration, PIC
simultaneously subtracts off the interference for each user produced
by the remaining ones. The MAI generated by other users is
reconstructed based on the tentative decisions from the previous
iteration. Therefore, the accuracy of the first iteration would
highly affect the PIC performance as error propagation occurs when
incorrect information imports. In this case, with the help of the
GL-PIC algorithm, the detection performance is improved. The key
novelty is that GL-PIC employs more reliable estimates by exploiting
prior knowledge of the constellation points. The proposed GL-PIC
algorithm is detailed in Table III.
\begin{table}[!htb]\scriptsize
\vspace{-1.5em}
\centering\caption{The GL-PIC algorithm}
\begin{tabular}{l}
\hline
$u_{k}(i)=\textbf{w}_{k}^{H}\textbf{y}_{sr_l}(i)$ \% soft outputs of all candidates\\
\textbf{for} k=1:K\\ \ \ \ \ \
\% Threshold comparison\\ \ \ \ \ \
\textbf{if} $u_{\textbf{t}_a(t)}(i)\notin \textbf{C}_{\rm grey}$\\ \ \ \ \ \ \ \ \ \
$\hat{b}_{\textbf{t}_a(t)}(i)=Q(u_{\textbf{t}_a(t)}(i))$\\ \ \ \ \ \
\textbf{else}\\ \ \ \ \ \ \ \ \
\textbf{prepared for reliability re-examination}\\ \ \ \ \ \
\textbf{end}\\
\textbf{end}\\
\textbf{Sort unreliable estimates $\textbf{t}_b$ in terms of the Euclidean distance} \\
\textbf{in a descending order}\\
\textbf{for} t=1:$n_q$ \% for the first $n_q$ most unreliable users\\ \ \ \ \ \
$\hat{b}_{\textbf{t}_b(t)}(i)=c^m$ \\
\textbf{end}\\
\textbf{for} t=$n_q$+1:length($\textbf{t}_b$)\\ \ \ \ \ \
$\hat{b}_{\textbf{t}_b(t)}(i)=Q(u_{\textbf{t}_b(t)}(i))$\\
\textbf{end}\\
\% Apply the ML rule to choose the best candidate list\\
$\textbf{b}^{\textrm{opt}}= \min _{\substack{1\leq j\leq N_c^{n_q}}}\parallel \textbf{y}_{sr_l}(i)-\textbf{H}_{sr_l}\textbf{b}^{j}(i)\parallel^2$\\
\% The three-iteration PIC process\\
\% $\textbf{b}^{\textrm{opt}}$ is used as the input \\
$\hat{b}_k^i=Q(\textbf{H}_{sr_l,k}^H\textbf{y}_{sr_l}-\sum\limits_{\substack{j=1\\j\neq k}}^{K}\textbf{H}_{sr_l,k}^H\textbf{H}_{sr_l,j}\hat{b}_j^{i-1})$\\
\hline
\vspace{-4.85em}
\end{tabular}
\end{table}

\section{proposed greedy multi-relay selection method}

In this section, a greedy multi-relay selection method is
introduced. For this problem, an exhaustive search of all possible
subsets of relays is needed to attain the optimum relay combination.
However, the major problem that prevents us from applying an
exhaustive search in practical communications is its very high
computational complexity. With $L$ relays involved in the
transmission, an exponential complexity of $2^L-1$ would be
required. This fact motivates us to seek alternative methods. By
eliminating the poorest relay-destination link stage by stage, the
standard greedy algorithm can be used in the selection process, yet
only a local optimum can be achieved. Unlike existing greedy
techniques, the proposed greedy multi-relay selection method can go
through a sufficient number of relay combinations and approach the
best one based on previous decisions. In the proposed relay
selection, the signal-to-interference-plus-noise ratio (SINR) is
used as the criterion to determine the optimum relay set. The
expression of the SINR for user $q$ is given by
\begin{equation}
 {\rm SINR_q} =\frac{E[|\textbf{w}_q^H\textbf{h}_q|^2]}{E[|\boldsymbol\eta|^2]+n},
\end{equation}
where $\textbf{w}_q$ denotes the RAKE receiver for user $q$, $E[|\boldsymbol\eta|^2]$
is the interference brought by all other users, and $n$ is the noise. For the RAKE receiver, the SINR for user $q$ is given by
\begin{equation}
{\rm SINR_q} =\frac{|\textbf{h}_q^H\textbf{h}_q|^2}
{\sum\limits_{\substack{k=1\\k\neq q}}^K|\textbf{h}_q^H
\textbf{h}_k|^2+\sigma_N^2\textbf{h}_q^H\textbf{h}_q},
\end{equation}
where $\textbf{h}_q$ is the channel vector for user $q$, $\textbf{H}$ is the channel matrix for all users. 
It should be mentioned that in various relay combinations, the channel vector $\textbf{h}_q$ for user $q$ $(q=1,2,...,K)$ is different as different relay-destination links are involved, $\sigma_N^2$ is the noise variance. This problem thus can be cast as the following optimization:
\begin{equation}
 {\rm SINR_{\Omega_{best}}}= \rm max \ \ \{\rm min({\rm SINR_{\Omega_{r(q)}}}), q=1,...,K\},
\end{equation}
where $\Omega_{r}$ denotes all possible combination sets $(r \leq L(L+1)/2)$ of any number of selected relays, ${\rm SINR_{\Omega_{r(q)}}}$ represents the SINR for user $q$ in set $\Omega_r$, min $({\rm SINR_{\Omega_{r(q)}}}) = {\rm SINR_{\Omega_r}}$ stands for the SINR for relay set $\Omega_r$ and
$\Omega_{\rm best}$ is the best relay set that provides the highest SINR.

\subsection{Standard greedy relay selection algorithm}
The standard greedy relay selection method works in stages by removing the single relay according to the channel path power, as given by
\begin{equation}
 P_{h_{r_ld}}=\textbf{h}_{r_ld}^H\textbf{h}_{r_ld},
\end{equation}
where $\textbf{h}_{r_ld}$ is the channel vector between the $l$-th relay and the destination. At the first stage, the initial SINR is determined
when all $L$ relays are involved in the transmission. Consequently, we cancel the worst relay-destination link and calculate the current SINR for the remaining $L-1$ relays, as compared with the previous SINR, if
\begin{equation}
{\rm SINR_{cur}}>{\rm SINR_{pre}}, \label{equation7}
\end{equation}
we update the previous SINR as
\begin{equation}
{\rm SINR_{pre}} = {\rm SINR_{cur}}, \label{equation8}
\end{equation}
and move to the third stage by removing the current poorest link and
repeating the above process. The algorithm stops either when ${\rm
SINR_{cur}}<{\rm SINR_{pre}}$ or when there is only one relay left.
The selection is performed once at the beginning of each packet
transmission.

\subsection{Proposed greedy relay selection algorithm}

In order to improve the performance of the standard algorithm, we
propose a new greedy relay selection algorithm that is able to
achieve a good balance between the performance and the complexity.
The proposed method differs from the standard technique as we drop
each of the relays in turns rather than drop them based on the
channel condition at each stage. The algorithm can be summarized as:
\begin{enumerate}
\item Initially, a set $\Omega_A$ that includes all $L$ relays is generated and its corresponding SINR is calculated, denoted by ${\rm SINR_{pre}}$.
\item For the second stage, we calculate the SINR for $L$ combination sets with each dropping one of the relays from $\Omega_A$. After that, we choose the combination set with the highest SINR for this stage, recorded as ${\rm SINR_{cur}}$.
\item Compare ${\rm SINR_{cur}}$ with the previous stage ${\rm SINR_{pre}}$, if (\ref{equation7}) is true, we save this corresponding relay combination as $\Omega_{\textrm{cur}}$ at this stage. Meanwhile, we update the ${\rm SINR_{pre}}$ as in (\ref{equation8}).
\item After moving to the third stage, we drop relays in turn again from $\Omega_{\textrm{cur}}$ obtained in stage two. $L-1$ new combination sets are generated, we then select the set with the highest SINR and repeat the above process in the following stages until either ${\rm SINR_{cur}}<{\rm SINR_{pre}}$ or there is only one relay left.
\end{enumerate}
This proposed greedy selection method considers the combination
effect of the channel condition so that additional useful sets are
examined. When compared with the standard greedy relay selection
method, the previous stage decision is more accurate and the global
optimum can be approached more closely. Furthermore, its complexity
is less than $L(L+1)/2$, which is much lower than the exhaustive
search. Similarly, the whole process is performed only once before
each packet and only needs to be repeated when the channels change.
The proposed greedy multi-relay selection algorithm is depicted in
Table IV.

\begin{table}[!htp]\scriptsize
\footnotesize
\centering\caption{The proposed greedy multi-relay selection algorithm}
\begin{tabular}{l}
\hline
$\Omega_A=[1,2,3,...L]$\% $\Omega_A$ denotes the set when all relays are involved\\
${\rm SINR_{\Omega_A}}=\textrm{min}({\rm SINR_{\Omega_{A(q)}}}), q=1,2,...K$\\
${\rm SINR_{pre}}={\rm SINR_{\Omega_A}}$\\
\textbf{for} stage =1 \textbf{to} $L-1$\\ \ \ \ \ \
\textbf{for} $r$=1 \textbf{to} $L+1$-stage\\ \ \ \ \ \ \ \ \ \
$\Omega_r=\Omega_A-\Omega_{A(r)}$\% drop each of the relays in turns\\ \ \ \ \ \ \ \ \ \
${\rm SINR_{\Omega_r}}=\textrm{min}({\rm SINR_{\Omega_r(q)}}), q=1,2,...,K$\\\ \ \ \ \
\textbf{end for}\\ \ \ \ \ \
${\rm SINR_{\textrm{cur}}}=\textrm{max}({\rm SINR_{\Omega_r}})$\\ \ \ \ \ \
$\Omega_{\textrm{cur}}=\Omega_{{\rm SINR_{cur}}}$\\ \ \ \ \ \
\textbf{if} ${\rm SINR_{cur}}>{\rm SINR_{pre}}$ \textbf{and} $|\Omega_{\textrm{cur}}|>1$\\ \ \ \ \ \ \ \ \ \
$\Omega_A=\Omega_{\textrm{cur}}$\\ \ \ \ \ \ \ \ \
${\rm SINR_{pre}}={\rm SINR_{cur}}$\\ \ \ \ \ \
\textbf{else}\\ \ \ \ \ \ \ \ \ \ \
\textbf{break}\\ \ \ \ \
\textbf{end if}\\
\textbf{end for}\\
\hline
\vspace{-4em}
\end{tabular}
\end{table}

\section{Analysis of the proposed algorithms}

In this section, we analyze the computational complexity required by
the proposed and existing interference cancellation algorithms and
the proposed greedy relay selection method.

\subsection{Computational complexity}

\begin{table}[!htbp]\scriptsize
\centering
\caption{Computational complexity of existing and proposed MUD algorithms}
\begin{tabular}{c|c}
\hline
\textbf{Algorithms} & \textbf{Computational Complexity (Flops)} \\
\cline{1-2}
\hline
\\
Matched filter      & $M(4L_p^2+4KL_p-2L_p+6K)-2K$ \\
\\
\hline
Conventional        & $M(4L_p^2+4KL_p-2L_p$  \\
SIC                 & $+18K-12)-4K+2$  \\

\hline
Conventional        & $M(4L_p^2+4KL_p-2L_p$ \\
PIC                 & $+10K+4K^2)-4K$ \\

\hline               & $8M^3+M^2(16K-8)$ \\
Linear MMSE receiver & $+M(4L_p^2+4KL_p-2L_p$\\
                     & $+4K+4)-2K$ \\

\hline              & $M(4L_p^2+4KL_p-2L_p+6K)$\\
Proposed GL-SIC     & $-2K+N_c^n(20MK-8Mn$ \\
                    & $+4M-2K+2n-2)$ \\

\hline              & $M(4L_p^2+4KL_p-2L_p$\\
Proposed GL-PIC     & $+10K+4K^2)-4K$ \\
                    & $+N_c^{n_q}(8MK+8M-2)$ \\
\hline
Standard Likelihood & $M(4L_p^2+4KL_p-2L_p-2K)$\\
(ML) detector       & $+N_c^K(8MK+8M-2)$\\
\hline
\end{tabular}\vspace{-3.5em}
\end{table}

We first compare the computational complexity of the proposed
(GL-SIC and GL-PIC) and other existing interference cancellation
algorithms in terms of the required floating point operations
(flops). The resulting complexity is calculated as a function of the
following parameters:

\begin{itemize}
\item Total number of users $K$.
\item The number of multipath channel components $L_p$.
\item The number of constellation points $N_c$ that correspond to the modulation type.
\item The parameter $M$ which corresponds to the length of the receive filters, where $M=N+L_p-1$ and $N$ is the spreading gain.
\end{itemize}

Specifically, in the GL-SIC algorithm, $n$ refers to the number of
users we considered per each stage, and in the GL-PIC algorithm,
$n_q$ represents the number of unreliable users that need to be
re-examined in the second processing stage. The required flops are
considered both in the case of real and complex matrix operations.
It is worth noting that, in real arithmetic, a multiplication
followed by an addition requires 2 flops while for the complex
numbers, 8 flops are required when an addition is executed after a
multiplication. As a result, it can be approximated that the
complexity of a complex matrix multiplication is 4 times of its real
counterpart.

Table. V illustrates a comparison of the computational complexity
for various existing detection methods and our proposed algorithms.
It is worth noting that the GL-SIC algorithm has variable complexity
according to different circumstances as an unpredictable number of
unreliable users may appear in any of the stages. As a result, the
corresponding worst-case scenario is evaluated when all $n$ users
are considered as unreliable at the first stage.

For each case shown in Table. V, the upper bound of the complexity
is given by the standard ML detector, where it explores all possible
combinations of the detected results and chooses the one with the
minimum Euclidean distance. However, when a large number of users
need to be considered, an exponential complexity growth would limit
its application in practical utilization. In contrast, with careful
control of the number of unreliable users $n$ and $n_q$ being
re-examined in both proposed algorithms, a substantial complexity
saving is achieved. Additionally, our proposed greedy list-based
algorithms offer a clear complexity advantage over the linear MMSE
receiver as they adopt the RAKE receiver as the front end, so that
the cubic complexity can be avoided. Another feature to highlight is
that although our proposed algorithms have a complexity slightly
higher than the matched filter, the conventional SIC and the
conventional PIC, they exhibit significant performance gains over
existing techniques.

\begin{figure}[!htb]
\begin{center}
\def\epsfsize#1#2{0.65\columnwidth}
\epsfbox{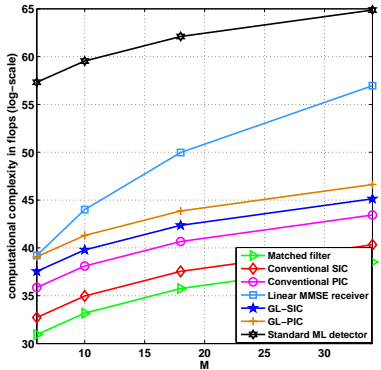} \caption{ Computational complexity in flops for
various MUD detectors} \label{fig3}
\end{center}
\end{figure}

In order to further investigate the computational complexity for various MUD techniques, we fix the number of users $K=10$, the number of multipath channel $L_p=3$ and assume the BPSK modulation is adopted. The required number of flops (log-scale)  of the proposed and existing MUD algorithms are simulated in Fig.\ref{fig3}, where in the GL-SIC algorithm, $n=2$ users are considered jointly at each stage and in the GL-PIC algorithm, $n_q=3$ unreliable users need to be re-examined in the second processing stage.
With the increase of the parameter $M$, the standard ML detector climbs significantly faster than other MUD schemes, which, from another point of view, demonstrates that the improvement in its performance is achieved at the expense of a large increase in computational complexity. A similar complexity trend for the linear MMSE receiver illustrated in Fig.\ref{fig3} shows a relatively lower complexity than the standard ML detector, however, its complexity still substantially exceeds that of the remaining strategies as a cubic cost is brought. Another important point observed in Fig.\ref{fig3} is that our proposed algorithms offer a moderately higher cost than the matched filter, the conventional SIC and the conventional PIC, whereas they provide a considerable performance advantage over these schemes, resulting in an attractive trade-off between complexity and performance.

\subsection{Greedy relay selection analysis}

The proposed greedy multi-relay selection method is a stepwise
backward selection algorithm, where we optimize the selection based
on the SINR criterion at each stage. We begin the process when all
relays participate in the transmission and then subtract off the
contributions brought by each of the relays from set of selected
relays of the previous stage. The relay combinations generated at
each stage are presented as follows:
\begin{equation*}
\begin{split}
&{\rm Stage \ 1:} \ \ \{\Omega_1^1\},\\
&{\rm Stage \ 2:} \ \ \{\Omega_1^2,\Omega_2^2,\Omega_3^2,...,\Omega_L^2\},\\
&\ \ \ \ \ \ \ \ \ \ \ \ \ \ \ \ \ \vdots \\
&{\rm Stage \ s:} \ \ \{\Omega_1^s,\Omega_2^s,\Omega_3^s,...,\Omega_{L+2-s}^s\},\\
&\ \ \ \ \ \ \ \ \ \ \ \ \ \ \ \ \ \vdots \\
&{\rm Stage \ L-1:} \ \ \{\Omega_1^{L-1},\Omega_2^{L-1},\Omega_3^{L-1} \},\\
&{\rm Stage \ L:} \ \ \{\Omega_1^L,\Omega_2^L\},
\end{split}
\end{equation*}
where $\Omega_i^s$ denotes the $i$-th relay combination at the $s$-th stage. Clearly, the maximum number of relay combinations that we have to consider for all $L$ stages is $1+2+3+...+L=(1+L)L/2$, since this algorithm stops either when ${\rm SINR_{cur}}<{\rm SINR_{pre}}$ or when there is only one relay left, the associated complexity for the proposed greedy relay selection strategy is less than $(1+L)L/2$.

Compared with the exhaustive search, which is considered as the optimum relay selection method, the number of relay combinations examined at each stage is given by
\begin{equation*}
\begin{split}
&{\rm Stage \ 1:} \ \ \{\Omega_1^1\},\\
&{\rm Stage \ 2:} \ \ \{\Omega_1^2,\Omega_2^2,\Omega_3^2,...,\Omega_L^2\},\\
&\ \ \ \ \ \ \ \ \ \ \ \ \ \ \ \ \ \ \ \ \ \ \vdots \\
&{\rm Stage \ s:} \ \ \{\Omega_1^s,\Omega_2^s,\Omega_3^s,...,\Omega_{\frac {L(L-1)...(L-s+2)}{(s-1)!}}^{s}\},\\
&\ \ \ \ \ \ \ \ \ \ \ \ \ \ \ \ \ \ \ \ \ \ \vdots \\
&{\rm Stage \ L-1:} \ \ \{\Omega_1^{L-1},\Omega_2^{L-1},\Omega_3^{L-1},...,\Omega_{\frac {L(L-1)}{2}}^{L-1} \},\\
&{\rm Stage \ L:} \ \ \{\Omega_1^L,\Omega_2^L\,\Omega_3^L,...,\Omega_L^L\}.
\end{split}
\end{equation*}
The total number of relay combinations can then be calculated as $C_L^L+C_L^{L-1}+C_L^{L-2}+...+C_L^2+C_L^1=2^L-1$, where each term $C_m^n=\frac{m(m-1)...(m-n+1)}{n!}$ represents the number of combinations that we choose, i.e., $n$ elements from $m$ elements$(m\geq n)$.
The proposed greedy algorithm provides a much lower cost with a moderate to large number of relays when compared with the exhaustive search as an exponential complexity is avoided.

In fact, the idea behind the proposed algorithm is to choose relay combinations in a greedy fashion. At each stage, we select the set of relays with the highest SINR and the previous stage result always affects the following stage set of relays candidates. Then we subtract off the contribution brought by each of the remaining relays and iterate on the residual. After several stages, the algorithm is able to identify the optimum relay set. To this end, we propose the following proposition. \\
\\
\textit{Proposition}: the proposed greedy algorithm achieves an SINR that is bounded as follows:
\begin{equation*}
\begin{split}
{\rm SINR_{\Omega standard}} \leq {\rm SINR_{\Omega proposed}} \leq {\rm SINR_{\Omega exhaustive}}
\end{split}
\end{equation*}
Proof:

From the proposed greedy algorithm, the set containing the selected relay at the $s$-th stage is given by\\
\begin{equation*}
\begin{split}
\Omega_{\rm proposed}^s=& \ \{m,n,...,p\}\\
=& \ {\rm max}\{\Omega_{\rm proposed}^{s-1}\setminus \Omega_{\rm proposed}^{s-1}(i),i\in[1,L+2-s]\},\\
\end{split}
\end{equation*}
where $\Omega^s\setminus \Omega^s(i)$ denotes a complementary set where we drop the $i$-th relay from the relay set $\Omega^s$. $m$, $n$ and $p$ represent the relay $m$, the relay $n$ and the relay $p$, respectively.

We first prove the lower bound for an arbitrary stage $s$ by induction, other stages can be obtained accordingly. Assuming both algorithms achieve the same set at stage $s$, we have
\begin{equation*}
\begin{split}
&\Omega_{\rm standard}^s= \ \{m,n,...,p\},\\
&\Omega_{\rm proposed}^s= \ \{m,n,...,p\},\\
\end{split}
\end{equation*}
which leads to the equality ${\rm SINR}_{\rm \Omega_{standard}^s}={\rm SINR}_{\rm \Omega_{proposed}^s}$, if we then proceed with the proposed greedy algorithm and choose a different set which provides a higher SINR, we have
\begin{equation*}
\begin{split}
&\Omega_{\rm standard}^s= \ \{m,n,...,p\},\\
&\Omega_{\rm proposed}^s= \ \{m,n,...,q\},\\
\end{split}
\end{equation*}
with the only different relay being $p \neq q$, and assuming that $q$ provides a higher SINR than $p$, we prove the inequality that
${\rm SINR_{\Omega standard}^s} \leq {\rm SINR_{\Omega proposed}^s}$.

We then investigate the upper bound by comparing the proposed algorithm and the exhaustive search at an arbitrary stage s, other stages can be obtained accordingly. At an arbitrary stage $s$, since $\Omega_{\rm proposed}^s$ is a candidate subset of the exhaustive search, we have
\begin{equation*}
\begin{split}
&\Omega_{\rm exhaustive}^s={\rm max} \ \{ \Omega_{\rm exhautive(i)}^{s}, i\in[1,C_L^{L+1-s}] \},\\
&\Omega_{\rm proposed}^s \in \{ \Omega_{\rm exhautive(i)}^{s}, i\in[1,C_L^{L+1-s}]\},\\
\end{split}
\end{equation*}
where $\Omega_{\rm exhaustive(i)}^s$ represents the $i$-th relay combination selected at the $s$-th stage of the exhaustive greedy relay selection method.

Assuming both strategies select the same relay combination at stage $s$, we have
\begin{equation*}
\begin{split}
&\Omega_{\rm proposed}^s= \ \{m,n,...,p\},\\
&\Omega_{\rm exhaustive}^s= \ \{m,n,...,p\},\\
\end{split}
\end{equation*}
this situation again leads to the equality that ${\rm SINR}_{\rm \Omega_{proposed}^s}={\rm SINR}_{\rm \Omega_{exhaustive}^s}$. In contrast, if the exhaustive search picks another relay set belongs to $\{ \Omega_{\rm exhautive(i)}^{s}, i\in[1,C_L^{L+1-s}] \}$ that provides a higher SINR, clearly, $\Omega_{\rm proposed}^s \neq \Omega_{\rm exhaustive}^s$, we can then obtain the inequality that ${\rm SINR}_{\Omega_{\rm proposed}^s} \leq {\rm SINR}_{\Omega_{\rm exhaustive}^s}$.

\section{Proposed cross-layer design}

In this section, we present and analyze a cross-layer design
strategy that combines the proposed MUD techniques with the proposed
greedy multi-relay selection algorithm for the uplink of the
cooperative DS-CDMA networks.  This approach jointly considers the
performance optimization across different layers of the network,
since inappropriate data detection and estimation that are executed
at the lower physical layer can spread incorrect information to the
data and link layer where relay selection strategy performs, causing
the loss of useful information and degradation of the overall system
performance. In this case, when improved data detection is obtained
at the physical layer, together with an effective relay selection, a
better system performance can be achieved.

As stated in previous sections, the system operates in two phases,
where for the first phase, the proposed MUD techniques are applied
and processed at each of the relays with a DF protocol, after the
detection process, the proposed greedy multi-relay selection
algorithm is then performed to seek the optimum relay combination.
In the second phase, the chosen relays take part in the transmission
in order to forward the information to the destination. After all
the data are received at the destination, the proposed MUD
algorithms are applied to recover the transmitted data.

Given the received data $\textbf{y}_{sd}$ and $\textbf{y}_{sr_l}$ at
the destination and each of the relays, we wish to optimize the
overall system performance in terms of the bit error rate (BER),
through the selection of the received signals $\textbf{y}_{rd}$ at
the destination from all relays, the accuracy of the detected
symbols $\hat{b}_{r_ld,k}$ at each of the relays and the detected
results $\hat{b}_k$ at the destination, subject to practical system
constraints ($K$, $L$, $L_p$, $b_k$, $\textbf{H}_{sd}$,
$\textbf{H}_{sr_l}$, $\textbf{H}_{r_ld}$, $\textbf{n}_{sd}$,
$\textbf{n}_{sr_l}$ and $\textbf{n}_{rd}$).
The proposed cross-layer design can be cast into the following optimization problem
\begin{equation} \label{equation20}
\begin{split}
\hspace{-0.5em} (\hat{\textbf{b}},\Omega^{\rm opt})=\min _{1\leq j\leq m} &\parallel
  \left[\hspace{-0.3em}
 \begin{array}{l}
   \ \ \ \ \ \textbf{H}_{sd} \textbf{b}\\
   \sum\limits_{l \in  \Omega^{s}} \textbf{H}_{r_ld} \hat{\textbf{b}}_{r_ld}\\
 \end{array} \hspace{-0.3em} \right] -
 \left[\hspace{-0.3em}
 \begin{array}{l}
   \ \ \ \ \ \textbf{H}_{sd} \textbf{b}^{j}\\
   \sum\limits_{l \in  \Omega^{s}} \textbf{H}_{r_ld} \textbf{b}^{j} \\
 \end{array}\hspace{-0.5em} \right]\parallel^2\\
 {\rm subject\ \ to} \ \ \ \ \ \ \ \ &\\
&m=N_c^{n_q} \textrm{or} \ N_c^n,\\
&\Omega^{\rm opt}= \Omega^s \ {\rm when} \ \ {\rm SINR}_{\Omega^s}< {\rm SINR}_{\Omega^{s-1}},\\
&{\rm SINR_{\Omega^s}}= {\rm max} \  \{ \ {\rm min} \ ({\rm SINR}_{\Omega^s_{i(k)}})\},\\
&k=1,2,...,K,\\
&s\leq L,\\
&i=1,2,...,L+2-s,
\end{split}
\end{equation}
where $\textbf{b}^{j}$ stands for the $j$-th candidate list
generated after applying the GL-SIC/GL-PIC algorithms at the
destination, $\Omega^s$ represents the selected relay combination at
the stage $s$, ${\rm SINR}_{\Omega_{i(k)}^s}$ is the SINR for the
$k$-th user in the $i$-th relay combination at stage $s$ and
$\Omega^{\rm opt}$ is the optimum relay combination obtained through
the proposed greedy relay selection method. The cross-layer
optimization in (\ref{equation20}) is a non-convex optimization
problem due to the discrete nature of the joint detection and relay
selection problems. We propose to solve it in two stages using the
proposed greedy detection and relay selection algorithms.

During the first phase, the received vector $\textbf{y}_{sr_l}$
passes through the proposed GL-SIC/GL-PIC algorithms at the relay
$l$, lists of candidate combinations $\textbf{b}^{j}_{r_ld}$ are
generated  in the lower physical layer and the corresponding
detected result $\hat{b}_{r_ld,k}$ is  then obtained via the
following ML selection
\begin{equation}
\hat{\textbf{b}}_{r_ld}= \min _{\substack{1\leq j\leq m, \textrm{where}\\m=N_c^{n_q} \textrm{or}\ N_c^n}}\parallel \textbf{y}_{sr_l}-\textbf{H}_{sr_l}\textbf{b}^{j}_{r_ld}\parallel^2.
\end{equation}
This interference cancellation operation affects the following
process in two different ways.
\begin{itemize}
\item The accuracy of $\hat{b}_{r_ld,k}$ directly controls the re-generated signals $\textbf{y}_{rd}$ received at the destination via the physical layer as
      can be verified from (\ref{equation0}), hence, it further affects the decisions $\hat{b}_k$ made at the end as (\ref{equation1}) computes.
\item Improper detection of $\hat{b}_{r_ld,k}$ can cause the error propagation spreads in the second phase.
\end{itemize}
Consequently, in the second phase, the proposed greedy relay
selection strategy is performed at the data and link layer, the
selection takes into account the physical layer characteristics as
appropriate detection result coming from the lower physical layer
can prevent error propagation spreading into the upper data and link
layer. In contrast, it also considers the features of the channel
combinations so that poor channels can be avoided.

In order to describe this process mathematically, we first define
the SINR for the $i$-th relay combination at an arbitrary stage $s$
as
\begin{equation}
{\rm SINR}_{\Omega_i^s} = \min \ \ \{ {\rm SINR}_{\Omega_{i(k)}^s}\}, k=1,2,...,K.
\end{equation}
This algorithm operates in stages, and the SINR for the selected relay combination at each stage is given by
\begin{equation*}
\begin{split}
{\rm SINR}_{\Omega^1}= & {\rm SINR}_{\Omega_A}, \Omega^1=\Omega_A=[1,2,3,...,L], \\
{\rm SINR}_{\Omega^2}= &\max \ \{ {\rm SINR}_{\Omega_i^2}\}, \Omega_i^2=\Omega^1\setminus \Omega^1(i),i=1,...,L,\\
{\rm SINR}_{\Omega^3}= &\max \ \{ {\rm SINR}_{\Omega_i^3}\}, \Omega_i^3=\Omega^2\setminus \Omega^2(i),i=1,...,L-1,\\
&\ \ \ \ \ \ \ \ \vdots \\
{\rm SINR}_{\Omega^L}= &\max  \ \{ {\rm SINR}_{\Omega_i^{L}}\},\Omega_i^L=\Omega^{L-1}\setminus \Omega^{L-1}(i), i=1,2.
\end{split}
\end{equation*}
The selection stops when ${\rm SINR}_{\Omega^s}< {\rm SINR}_{\Omega^{s-1}}$ is achieved, and the optimum relay combination is then computed as
$\Omega^{\rm opt}=\Omega^s$. After that, the selected relays continue to forward the re-generated signals to the destination in the second phase.

At the destination, after we receive both the signals from the
direct links and the selected relays, we then apply the
GL-SIC/GL-PIC algorithms again to obtain lists of candidates
combinations $\textbf{b}^{j}$, and the ML algorithm is adopted
afterwards to choose the optimum detection list as given by
\begin{equation}
\begin{split}
\hspace{-0.5em} \hat{\textbf{b}}&= \min _{\substack{1\leq j\leq m, \textrm{where}\\m=N_c^{n_q} \textrm{or}\ N_c^n}}\parallel
\left[\hspace{-0.3em}
 \begin{array}{l}
   \ \ \ \ \ \textbf{y}_{sd}\\
   \sum\limits_{l \in  \Omega^{\rm opt}} \textbf{y}_{r_ld}\\
 \end{array} \hspace{-0.3em} \right] -
\left[\hspace{-0.3em}
 \begin{array}{l}
   \ \ \ \ \ \textbf{H}_{sd}\\
   \sum\limits_{l \in  \Omega^{\rm opt}} \textbf{H}_{r_ld}\\
 \end{array}\hspace{-0.3em} \right]
\hspace{-0.5em}
 \begin{array}{l}
   \textbf{b}^{j}\\
 \end{array}\parallel^2\\
 &=\min _{\substack{1\leq j\leq m, \textrm{where}\\m=N_c^{n_q} \textrm{or}\ N_c^n}}\parallel
 \left[\hspace{-0.3em}
 \begin{array}{l}
   \ \ \ \ \ \textbf{H}_{sd} \textbf{b}\\
   \sum\limits_{l \in  \Omega^{\rm opt}} \textbf{H}_{r_ld} \hat{\textbf{b}}_{r_ld}\\
 \end{array} \hspace{-0.3em} \right] -
 \left[\hspace{-0.3em}
 \begin{array}{l}
   \ \ \ \ \ \textbf{H}_{sd} \textbf{b}^{j}\\
   \sum\limits_{l \in  \Omega^{\rm opt}} \textbf{H}_{r_ld} \textbf{b}^{j} \\
 \end{array}\hspace{-0.5em} \right]\parallel^2.\\
\end{split}
\end{equation}

The proposed cross-layer design is detailed in Table VI.
\begin{table}[!htb]\scriptsize
\centering\caption{The cross-layer desgin}
\begin{tabular}{l}
\hline
\textbf{Phase I}\\
\%received signals from the source-destination link\\
$\textbf{y}_{sd}=\textbf{H}_{sd}\textbf{b}_k$\\
\%received signals from the source to the $l$-th relay\\
$\textbf{y}_{sr_l}=\textbf{H}_{sr_l}\textbf{b}_k$\\
\% Interference cancellation process at each of the relays\\
\textbf{Apply the GL-SIC/GL-PIC algorithms }\\
\textbf{at each of the relays to obtain} $\textbf{b}^{j}_{r_ld}$\\
\% Apply the ML rule to select $\hat{\textbf{b}}_{r_ld}$ from $\textbf{b}^{j}_{r_ld}$\\
$\hat{\textbf{b}}_{r_ld}= \min _{\substack{1\leq j\leq m, \textrm{where}\\m=N_c^{n_q} \textrm{or}\ N_c^n}}\parallel \textbf{y}_{sr_l}-\textbf{H}_{sr_l}\textbf{b}^{j}_{r_ld}\parallel^2$\\
\textbf{Phase II}\\
\textbf{Apply the greedy multi-relay selection method}\\
${\rm SINR}_{\Omega_i^s} = \min \ \ \{ {\rm SINR}_{\Omega_{i(k)}^s}\}, k=1,2,...,K$\\
${\rm SINR}_{\Omega^{s}}= \max \ \{ {\rm SINR}_{\Omega_i^{s}}\}, \Omega_i^{s}=\Omega^{s-1}\setminus \Omega^{s-1}(i)$,\\\ \ \ \ \ \ \ \ \ \ \ \ \ \
$i=1,2,...,L+2-s$\\
$\Omega^{\rm opt}= \Omega^s \ {\rm when} \ \ {\rm SINR}_{\Omega^s}< {\rm SINR}_{\Omega^{s-1}}$\\
\%received signals from the selected relays to the destination\\
$\textbf{y}_{rd}=\sum\limits_{l \in  \Omega^{\rm opt}} \textbf{H}_{r_ld} \hat{\textbf{b}}_{r_ld}$\\
\textbf{Apply the GL-SIC/GL-PIC algorithms}\\
\textbf{at the destination to obtain} $\textbf{b}^{j}$\\
\% Apply the ML rule to select $\hat{\textbf{b}}_k$ from $\textbf{b}^{j}$\\
\begin{minipage}{3.18in}
\begin{equation*}
\begin{aligned}
\begin{split}
\hspace{-0.75em} \hat{\textbf{b}}= \min _{\substack{1\leq j\leq m, \textrm{where}\\m=N_c^{n_q} \textrm{or}\ N_c^n}}\parallel
 \left[\hspace{-0.3em}
 \begin{array}{l}
   \ \ \ \ \ \textbf{H}_{sd} \textbf{b}\\
   \sum\limits_{l \in  \Omega^{\rm opt}} \textbf{H}_{r_ld} \hat{\textbf{b}}_{r_ld}\\
 \end{array} \hspace{-0.3em} \right] -
 \left[\hspace{-0.3em}
 \begin{array}{l}
   \ \ \ \ \ \textbf{H}_{sd} \textbf{b}^{j}\\
   \sum\limits_{l \in  \Omega^{\rm opt}} \textbf{H}_{r_ld} \textbf{b}^{j}\\
 \end{array}\hspace{-0.5em} \right]\parallel^2.\\
\end{split}
\end{aligned}
\end{equation*}
\end{minipage}\\
\hline
\vspace{-5.85em}
\end{tabular}
\end{table}

\section{Simulations}

In this section, a simulation study of the proposed multiuser
detectors and the low cost greedy multi-relay selection method is
carried out. The DS-CDMA network uses randomly generated spreading
codes of length $N=32$ and $N=16$, it also employs $L_p=3$
independent paths with the power profile $[0\rm dB,-3\rm dB,-6\rm
dB]$ for the transmission link. The corresponding channel
coefficients are taken as uniformly random variables and normalized
to ensure the total power is unity. We assume perfectly known
channels at the receiver. Equal power allocation with normalization
is assumed to guarantee no extra power is introduced during the
transmission.  The grey area in the GL-SIC and GL-PIC algorithm is
determined by the threshold where $d_{th}=0.25$. We consider packets
with 1000 BPSK symbols and average the curves over 300 trials. For
the purpose of simplicity, $n=2$ users are considered in the GL-SIC
scheme at each stage and for the GL-PIC strategy, a three-iteration
PIC process is adopted. The following simulations are compared and
analyzed in both non-cooperative and cooperative scenarios.

\begin{figure}[!htb]
\begin{center}
\def\epsfsize#1#2{0.65\columnwidth}
\epsfbox{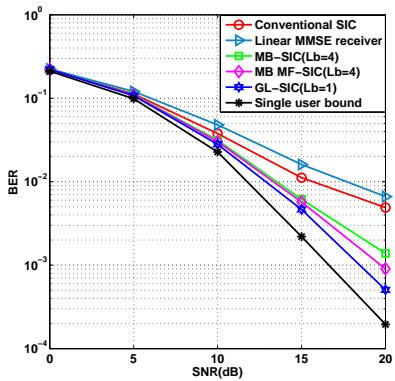} \caption{ GL-SIC comparison in non-cooperative
system with 20 users over Rayleigh fading channel}  \label{fig4}
\end{center}
\end{figure}

The first example shown in Fig.\ref{fig4} illustrates the
performance comparison between the proposed GL-SIC interference
suppression technique and other multiuser detection methods over the
Rayleigh fading channel. The proposed GL-SIC algorithm uses the
spreading codes with length $N=32$ and the overall system is
equipped with 20 users that only takes into account the source to
the destination link. The conventional SIC detector is the standard
SIC with RAKE receivers employed at each stage and the Multi-branch
Multi-feedback SIC (MB MF-SIC) detection algorithm mentioned in
\cite{Li1} is presented here for comparison purposes. We also
produce the simulation results for the multi-branch SIC (MB-SIC)
detector where four parallel branches with different detection
orders are employed. Specifically, the detection order for the first
branch is obtained through a power decreasing level, while the
detection orders for the remaining three are attained by cyclically
shifting the order index from the previous branch to right by one
position, similarly, RAKE receivers are adopted at each cancellation
stage. Simulation results reveal that our proposed single branch
GL-SIC significantly outperforms the linear MMSE receiver, the
conventional SIC and exceeds the performance of MB-SIC with $L_b=4$
and MB MF-SIC with $L_b=4$ for the same BER performance.

%

\begin{figure}[!htp]
\begin{center}
\def\epsfsize#1#2{0.65\columnwidth}
\epsfbox{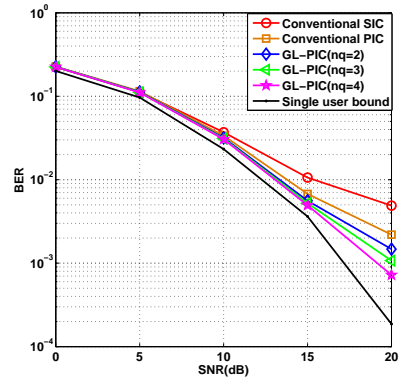} \caption{ GL-PIC comparison in non-cooperative
system with 20 users over Rayleigh fading channel}  \label{fig5}
\end{center}
\end{figure}

In the second example, the BER performance of the analyzed detection
schemes is then examined for the proposed GL-PIC detector employed
in the direct transmission over the Rayleigh fading channel, $N=32$
and the user number is 20. As depicted in Fig.\ref{fig5}, the
results compare the BER versus SNR performance between the
conventional detectors and the GL-PIC techniques with different
number of unreliable users being re-examined, the figure advises
that the GL-PIC algorithm performs better than the conventional SIC
detector and the conventional PIC detectors, both with RAKE
receivers employed at each cancellation stage. Moreover, with the
additional number of unreliable users being re-examined, extra
performance gains can be obtained. However, in this non-cooperative
Rayleigh fading system, the performance improvement is slight and
the detection capability is not that good when compared with the
GL-SIC scheme.

The next scenario illustrated in Fig. \ref{fig6}(a) shows the BER
versus SNR plot for the cross-layer design using the proposed
detectors and the greedy relay selection method, where we apply the
GL-SIC/GL-PIC algorithms at both the relays and the destination in
an uplink cooperative scenario with 10 users, 6 relays and spreading
gain $N=16$. The performance bounds for an exhaustive search of
different detectors are presented here for comparison purposes,
where it examines all possible relay combinations and picks the best
one with the highest SINR. From the results, it can be seen that
with the relay selection, the GL-SIC $(L_b=1)$ detector performs
better than the GL-PIC detector in high SNR region. Furthermore, the
BER performance curves of our proposed relay selection algorithm
approach almost the same level of the exhaustive search, whilst
keeping the complexity reasonably low for practical utilization.

In contrast, when the algorithms are assessed in terms of BER versus
number of users in Fig.\ref{fig6}(b) with a fixed SNR=15dB.
Similarly, we apply both the GL-SIC and the GL-PIC detectors at both
the relays and destination. The results indicate that the overall
system performance degrades as the number of users increases. In
particular, this figure also suggests that our proposed greedy relay
selection method has a big advantage for situations without a high
load and can approaches the exhaustive search very closely with a
relatively lower complexity. Additionally, the BER performance
curves of GL-SIC detector is better than the GL-PIC detector
especially for a large number of users.

\begin{figure}[!htb]
\begin{center}
\def\epsfsize#1#2{0.65\columnwidth}
\epsfbox{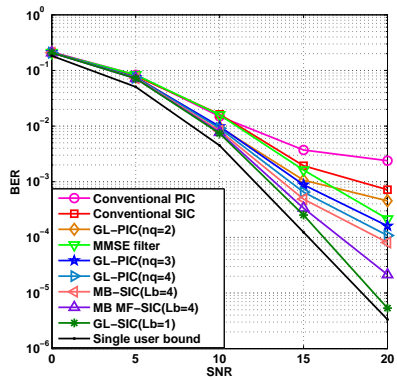} \caption{ BER versus SNR for uplink cooperative
system with different filters employed at the relays and the
destination}  \label{fig7}
\end{center}
\end{figure}

In order to further verify the performance for the proposed cross-layer design, we compare the effect of different detectors with 10 users and 6 relays when this new greedy multi-relay selection algorithm is applied in the system. The results depicted in Fig.\ref{fig7} with spreading gain $N=16$ indicate that the GL-SIC $(L_b=1)$ approach allows a more effective reduction of BER and achieves the best performance that is quite close to the single user scenario, followed by the MB MF-SIC $(L_b=4)$ detector, the MB-SIC $(L_b=4)$ detector, the GL-PIC detector, the linear MMSE receiver, the conventional SIC detector and the conventional PIC detector. Additionally, it is worth noting that some extra performance gains are attained for the GL-PIC approach as more $n_q$ unreliable users are selected and re-examined.

\section{Conclusions}

In this work, we have presented the GL-SIC and GL-PIC interference
cancellation algorithms, which can approach the ML performance at a
much lower cost than competing techniques. We have also proposed a
greedy multi-relay selection algorithms that outperforms existing
greedy algorithms and obtains a performance close to an exhaustive
search. A novel cross-layer design strategy that incorporates GL-SIC
or GL-PIC, and a greedy multi-relay selection algorithm for the
uplink of cooperative DS-CDMA systems has been also presented. This
approach effectively reduces the error propagation generated at the
relays, avoiding the poorest relay-destination link while requiring
a low complexity. Simulation results demonstrate that the
performance of the proposed cross-layer design is superior to
existing techniques, can approach an interference-free scenario and
be applied to other wireless systems.

\bibliographystyle{IEEEtran}
\bibliography{reference}

\begin{thebibliography}{10}
\providecommand{\url}[1]{#1}
\csname url@samestyle\endcsname
\providecommand{\newblock}{\relax}
\providecommand{\bibinfo}[2]{#2}
\providecommand{\BIBentrySTDinterwordspacing}{\spaceskip=0pt\relax}
\providecommand{\BIBentryALTinterwordstretchfactor}{4}
\providecommand{\BIBentryALTinterwordspacing}{\spaceskip=\fontdimen2\font plus
\BIBentryALTinterwordstretchfactor\fontdimen3\font minus
  \fontdimen4\font\relax}
\providecommand{\BIBforeignlanguage}[2]{{%
\expandafter\ifx\csname l@#1\endcsname\relax
\typeout{** WARNING: IEEEtran.bst: No hyphenation pattern has been}%
\typeout{** loaded for the language `#1'. Using the pattern for}%
\typeout{** the default language instead.}%
\else
\language=\csname l@#1\endcsname
\fi
#2}}
\providecommand{\BIBdecl}{\relax}
\BIBdecl

\bibitem{Proakis}
J.G.Proakis, \emph{Digital Communications}, 4th~ed.\hskip 1em plus 0.5em minus
  0.4em\relax New York, USA: McGraw-Hill, Inc, 2011.

\bibitem{sendonaris}
A.~Sendonaris, E.~Erkip, and B.~Aazhang, ``User cooperation diversity - parts
  {I} and {II},'' \emph{IEEE Trans. Communication.}, vol.~51, no.~11, pp.
  1927--1948, November 2003.

\bibitem{Venturino}
L.~Venturino, X.~Wang, and M.~Lops, ``Multi-user detection for cooperative
  networks and performance analysis,'' \emph{IEEE Trans. Signal Processing},
  vol.~54, no.~9, pp. 3315--3329, September 2006.

\bibitem{laneman04}
J.~N. Laneman and G.~W. Wornell, ``Cooperative diversity in wireless networks:
  Efficient protocols and outage behaviour,'' \emph{IEEE Trans. Inf. Theory},
  vol.~50, no.~12, pp. 3062--3080, December 2004.

\bibitem{Bai}
L.~Bai, L.~Zhao, and Z.~liao, ``A novel cooperation scheme in wireless sensor
  networks,'' \emph{IEEE Wireless Communications and Networking Conference},
  pp. 1889--1893, Las Vegas, NV, Apr. 2008.

\bibitem{Souryal}
M.~R. Souryal, B.~R. Vojcic, and R.~L. Pickholtz, ``Adaptive modulation in ad
  hoc {DS/CDMA} packet radio networks,'' \emph{IEEE Trans. Communication},
  vol.~54, no.~4, pp. 714--725, Apr. 2006.

\bibitem{Levorato}
M.~Levorato, S.~Tomasin, and M.~Zorzi, ``Cooperative spatial multiplexing for
  ad hoc networks with hybrid {ARQ}: System design and performance analysis,''
  \emph{IEEE Trans. Communication}, vol.~56, no.~9, pp. 1545--1555, Sep. 2008.

\bibitem{Verdu1}
S.Verdu, \emph{Multiuser Detection}.\hskip 1em plus 0.5em minus 0.4em\relax
  Cambridge, 1998.

\bibitem{Verdu2}
S.~Verdu, ``Minimum probability of error for asynchronous gaussian
  multiple-access channels,'' \emph{IEEE Trans. Inform. Theory}, vol. IT32,
  no.~1, pp. 85--96, Jan. 1986.

\bibitem{Lupas}
R.~Lupas and S.~Verdu, ``Linear multiuser detectors for synchronous
  code-division multiple-access channels,'' \emph{IEEE Trans. Inform. Theory},
  vol.~35, no.~1, pp. 123--136, Jan. 1989.

\bibitem{Patel}
P.~Patel and J.~Holtzman, ``Analysis of a simple successive interference
  cancellation scheme in {DS/CDMA} systems,'' \emph{IEEE J. Select. Areas
  Commun.}, vol.~12, no.~5, pp. 796--807, Jun. 1994.

\bibitem{Varanasi}
M.~K. Varanasi and B.~Aazhang, ``Multistage detection in asynchronous
  code-division multiple-access communications,'' \emph{IEEE Trans.
  Communication}, vol.~38, no.~4, pp. 509--519, Apr. 1990.

\bibitem{RCDL1}
R.~C. de~Lamare and R.~Sampaio-Neto, ``Minimum mean-squared error iterative
  successive parallel arbitrated decision feedback detectors for {DS-CDMA}
  systems,'' \emph{IEEE Trans. Communication}, vol.~56, no.~5, pp. 778--789,
  May. 2008.

\bibitem{Jing}
Y.~Jing and H.~Jafarkhani, ``Single and multiple relay selection schemes and
  their achievable diversity orders,'' \emph{IEEE Trans. Wireless Commun.},
  vol.~8, no.~3, pp. 1084--1098, Mar 2009.

\bibitem{Clarke}
P.~Clarke and R.~C. de~Lamare, ``Transmit diversity and relay selection
  algorithms for multi-relay cooperative {MIMO} systems,'' \emph{IEEE Trans.
  Veh. Technol}, vol.~61, no.~3, pp. 1084--1098, Mar 2012.

\bibitem{Ding}
M.~Ding, S.~Liu, H.~Luo, and W.~Chen, ``{MMSE} based greedy antenna selection
  scheme for {AF} {MIMO} relay systems,'' \emph{IEEE Signal Process. Lett},
  vol.~17, no.~5, pp. 433--436, May 2010.

\bibitem{Song}
S.~Song and W.~Chen, ``{MMSE} based greedy eigenmode selection for {AF} {MIMO}
  relay channels,'' \emph{IEEE Globecom}, Anaheim, CA, Dec. 2012.

\bibitem{Talwar}
S.~Talwar, Y.~Jing, and S.~Shahbazpanahi, ``Joint relay selection and power
  allocation for two-way relay networks,'' \emph{IEEE Signal Process. Lett},
  vol.~18, no.~2, pp. 91--94, Feb 2011.

\bibitem{Tropp}
J.~Tropp, ``Greedy is good: Algorithmic results for sparse approximation,''
  \emph{IEEE Trans. Inf. Theory}, vol.~50, no.~10, pp. 2231--2242, Oct. 2004.

\bibitem{Flury}
R.~Flury, S.~V. Pemmaraju, and R.~Wattenhofer, ``Greedy routing with bounded
  stretch,'' \emph{IEEE Infocom.}, Rio de Janeiro, Brazil, Apr. 2009.

\bibitem{Jia}
Y.~Jia, E.~Yang, D.~He, and S.~Chan, ``A greedy re-normalization method for
  arithmetic coding,'' \emph{IEEE Trans. Communication}, vol.~55, no.~8, pp.
  1494--503, Aug. 2007.

\bibitem{RCDL2}
R.~C. de~Lamare, ``Joint iterative power allocation and linear interference
  suppression algorithms for cooperative {DS-CDMA} networks,'' \emph{IET,
  Communications}, vol.~6, no.~13, pp. 1930--1942, Sep. 2012.

\bibitem{Chen}
W.~Chen, L.~Dai, K.~B. Letaief, and Z.~Cao, ``A unified cross-layer framework
  for resource allocation in cooperative networks,'' \emph{IEEE Trans. Wireless
  Commun.}, vol.~7, no.~8, pp. 3000--3012, Aug. 2008.

\bibitem{Cao}
Y.~Cao and B.~Vojcic, ``{MMSE} multiuser detection for cooperative diversity
  {CDMA} systems,'' \emph{IEEE Wireless Communications and Networking
  Conference}, pp. 42--47, Atlanta, GA, Apr. March.

\bibitem{int}
R.~C. de~Lamare and R.~C. de~Lamare, ``Adaptive reduced-rank mmse filtering
  with interpolated fir filters and adaptive interpolators,'' \emph{IEEE Signal
  Processing Letters}, vol.~12, no.~3, March 2005.

\bibitem{Meng}
R.~Meng, R.~C. de~Lamare, and V.~H. Nascimento, ``Sparsity-aware affine
  projection adaptive algorithms for system identification,'' \emph{in Proc.
  Sensor Signal Processing for Defence Conference}, London, UK, 2011.

\bibitem{l1cg}
Z.~Yang, R.~de~Lamare, and X.~Li, ``Sparsity-aware space-time adaptive
  processing algorithms with l1-norm regularisation for airborne radar,''
  \emph{Signal Processing, IET}, vol.~6, no.~5, pp. 413--423, July 2012.

\bibitem{zhaocheng}
------, ``L1-regularized stap algorithms with a generalized sidelobe canceler
  architecture for airborne radar,'' \emph{Signal Processing, IEEE Transactions
  on}, vol.~60, no.~2, pp. 674--686, Feb 2012.

\bibitem{alt}
R.~C. de~Lamare and R.~C. de~Lamare, ``Sparsity-aware adaptive algorithms based
  on alternating optimization and shrinkage,'' \emph{IEEE Signal Processing
  Letters}, vol.~21, no.~2, pp. 225--229, January 2014.

\bibitem{jiolms}
R.~C. de~Lamare and R.~Sampaio-Neto, ``Reduced--rank adaptive filtering based
  on joint iterative optimization of adaptive filters,'' \emph{IEEE Signal
  Process. Lett.}, vol.~14, no.~12, pp. 980--983, December 2007.

\bibitem{jiols}
------, ``Reduced-rank space--time adaptive interference suppression with joint
  iterative least squares algorithms for spread-spectrum systems,'' \emph{IEEE
  Transactions Vehicular Technology}, vol.~59, no.~3, pp. 1217--1228, March
  2010.

\bibitem{jiomimo}
------, ``Adaptive reduced-rank equalization algorithms based on alternating
  optimization design techniques for {MIMO} systems,'' \emph{IEEE Transactions
  on Vehicular Technology}, vol.~60, no.~6, pp. 2482--2494, July 2011.

\bibitem{jidf}
------, ``Adaptive reduced-rank processing based on joint and iterative
  interpolation, decimation, and filtering,'' \emph{IEEE Transactions on Signal
  Processing}, vol.~57, no.~7, pp. 2503--2514, July 2009.

\bibitem{fa10}
R.~Fa, R.~C. de~Lamare, and L.~Wang, ``Reduced-rank stap schemes for airborne
  radar based on switched joint interpolation, decimation and filtering
  algorithm,'' \emph{IEEE Transactions on Signal Processing}, vol.~58, no.~8,
  pp. 4182--4194, August 2010.

\bibitem{saabf}
S.~Li, R.~C. de~Lamare, and R.~Fa, ``Reduced-rank linear interference
  suppression for ds-uwb systems based on switched approximations of adaptive
  basis functions,'' \emph{IEEE Transactions on Vehicular Technology}, vol.~60,
  no.~2, pp. 485--497, Feb 2011.

\bibitem{barc}
R.~C. de~Lamare, R.~Sampaio-Neto, and M.~Haardt, ``Blind adaptive constrained
  constant-modulus reduced-rank interference suppression algorithms based on
  interpolation and switched decimation,'' \emph{IEEE Transactions on Signal
  Processing}, vol.~59, no.~2, pp. 681--695, Feb 2011.

\bibitem{honig}
M.~L. Honig and J.~S. Goldstein, ``Adaptive reduced-rank interference
  suppression based on the multistage wiener filter,'' \emph{IEEE Transactions
  on Communications}, vol.~50, no.~6, June 2002.

\bibitem{mswfccm}
R.~C. de~Lamare, M.~Haardt, and R.~Sampaio-Neto, ``Blind adaptive constrained
  reduced-rank parameter estimation based on constant modulus design for cdma
  interference suppression,'' \emph{IEEE Transactions on Signal Processing},
  vol.~56, no.~6, June 2008.

\bibitem{locsme}
H.~Ruan and R.~de~Lamare, ``Robust adaptive beamforming using a low-complexity
  shrinkage-based mismatch estimation algorithm,'' \emph{Signal Processing
  Letters, IEEE}, vol.~21, no.~1, pp. 60--64, Jan 2014.

\bibitem{Li1}
P.~Li and R.~C. de~Lamare, ``Multiple feedback successive interference
  cancellation detection for multiuser {MIMO} systems,'' \emph{IEEE Trans.
  Wireless Commun.}, vol.~10, no.~8, pp. 2434--2439, Aug. 2011.

\bibitem{Li2}
P.~Li, R.~C. de~Lamare, and R.~Fa, ``Multi-feedback successive interference
  cancellation with multi-branch processing for {MIMO} systems,''
  \emph{Vehicular Technology Conference (VTC Spring)}, pp. 1--5, May. 2011.

\end{thebibliography}
\end{document}